\documentclass[a4paper]{jpconf}
\usepackage{graphicx}
\begin{document}
\title{OAUNI astronomical photometry: Stellar variability of FO Aqr on 2016 low state}

\author{A Pereyra$^1$$^,$$^2$}

\address{$^1$ Instituto Geof\'isico del Per\'u, \'Area Astronom\'ia, Badajoz 169, Ate, Lima, Per\'u}
\address{$^2$ Faculty of Science, National University of Enginnering, Av.T\'upac Amaru 210, Lima, Per\'u}

\ead{apereyra@igp.gob.pe}

\begin{abstract}
This work reports a photometric monitoring of the intermediate polar cataclysmic variable star FO Aqr. This is part of the ongoing OAUNI stellar variability program. Around 1200 individual measurements (or $\sim$ 9 hrs of observations) were gathered distributed in five nights. The observation epoch was coincident with the 2016 low state brightness of FO Aqr. Good quality light curves detect unambiguously the expected stellar variability of FO~Aqr. The analysis using periodograms let to determine the main two known periods, P$_{1}$~=~20.380~$\pm$~.003~min and P$_{2}$~=~11.076~$\pm$~.001~min. Our full five-nights analysis shows that P$_{1}$, the spin white dwarf period, is more prominent than P$_{2}$, the one-half of the beat period between the spin and orbital periods. Nevertheless, individual light curves show than, at least in one case, a reverse pattern is found. It suggests a dependence of the beat period with the orbital phase of the system with important variations at the same phase interval.

\end{abstract}

\section{Introduction}
Cataclysmic variables (CV) are interacting binary stars with transferring mass of a normal star to its companion. This process yields occasional energetic outburst events with a strong mutual influence between their components. The canonical model is represented by a white dwarf star that is accreting material from a lower mass and very close red star with an accretion disk formed around the first one \cite{con07}.

An Intermediate Polar is a CV with a white dwarf and a cool main-sequence secondary star. In these systems the canonical model requires that the inner disk is disrupted by the magnetic field of the rapidly rotating white dwarf with rotational periods range from tens of seconds to few hours \cite{pat94}. Approximately 1\% of the known CVs are confirmed as intermediate polar stars \cite{dow06}.

FO Aqr is an intermediary polar with a system orbiting period of 4.85 hours \cite{osb89}. The system underwent a low state in 2016 (decrease of the optical luminosity by 2 mag, from a normal state with V = 13.5 to a faint state with V = 15.5), from which it recovered slowly and steadily over a time scale of several months \cite{lit16a}. The system has a very strong optical pulsation of 20.9 minutes, associated with the spin period of the accreting white dwarf (WD) \cite{pat83}. In the 2016 lowest state, a very prominent 11.26 minute was claimed \cite{lit16b}, corresponding with one-half of the beat period between the spin and orbital periods. In normal state, this last period is comparatively weak \cite{ken16}. After this episode the system come back slowly to its normal high state brightness, indicating a temporary drop-off in the mass-transfer rate between the two stars \cite{lit16a}.

The astronomical observatory of the National University of Enginnering \cite{per15} (OAUNI in spanish) is operating since 2015 at the Huancayo site of the Geophysical Institute of Peru. The observation present here are part of the cataclysmic variables OAUNI program.

\section{Observations and Reductions}
The FO Aqr observations were gathered using the 0.5m OAUNI telescope \cite{per15}. The measurements were performed during one night on July 2016 and four nights on August 2016. The used detector was a front-illuminated CCD STXL-6303E (manufactured by SBIG company) of 3072$\times$2048 pixels$^2$ and 9$\mu$m/pixel. This detector along with the focal ratio f/8.2 of the optical system yields a plate scale of 0.45$"$/pixel and a field-of-view (FOV) of 23$'\times$15$'$. A \textit{R} broadband filter was used in all the observations with individual integration times of 20 sec in each sequence. Table \ref{foaqrtab} shows a log of FO Aqr observations. The observation date is indicated in Col. (1) with the number of measurements in a given observation sequence in Col (2). Around 1200 individual measurements in total were gathered. The time interval (in hours) including overheads for each sequence is indicated in Col. (3). Considering all the acquired sequences the total FO Aqr monitoring was of 8.9 hours. Col. (4) and (5) show the mean seeing and airmass in each sequence. In general, the night conditions were stable including a couple night with good seeing. All the measurements were done with the target well positioned on the sky (airmass below than 1.5).

\begin{table}
\caption{\label{foaqrtab}FO Aqr observation log.}
\begin{center}
\small
\begin{tabular}{lcccccc}
\br
Date & \textit{N} & interval & $<$seeing$>$ & $<$airmass$>$ & \textit{R}$^{a}$ & $\phi_{int}$ \\
     &            & (hrs)    &  ($"$)       &               &  (mag)            &         \\
  (1)		&  	(2)         & (3)       & (4)    &   (5)    &  (6)  & (7) \\		    
\mr
2016/07/14 & 180 & 1.3  & 1.61  &  1.33  &   15.23 (0.11) &  0.72$-$1.00  \\
2016/08/02 & 300 & 2.2  & 2.27  & 1.19   &   15.06 (0.14) &  0.37$-$0.83  \\
2016/08/03 & 320 & 2.4  & 2.10  & 1.19   &   15.04 (0.12) &  0.31$-$0.80  \\
2016/08/04 & 300 & 2.2  & 1.68  & 1.42   &   14.86 (0.19) &  0.21$-$0.67  \\
2016/08/05 & 91  & 0.7  & 2.03  & 1.30   &   15.06 (0.11) &  0.36$-$0.50  \\
\br
$^{a}$ errors in parenthesis. 
\end{tabular}
\end{center}
\end{table}

\begin{figure}
\begin{minipage}{8.0cm}
\includegraphics[width=8.0cm]{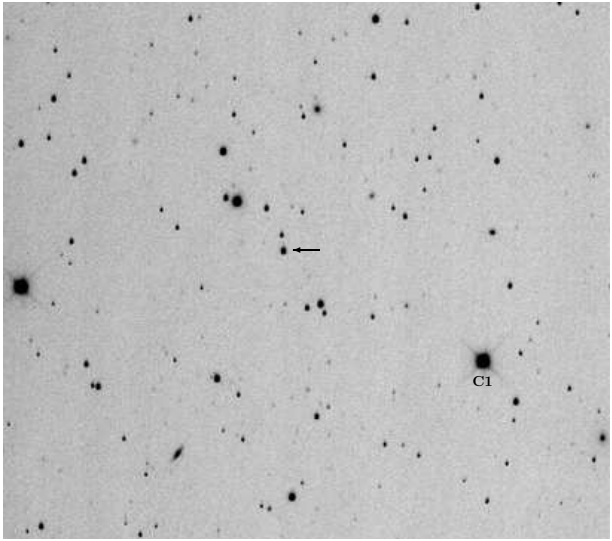}
\end{minipage}
\hspace{0.3cm}
\begin{minipage}[t]{7.7cm}
\caption{\label{foaqrfov}OAUNI stacked image (300 images$\times$20s = 2.2h, including overhead) of FO Aqr observed on 2016/08/02. North is top and East is left. The FOV section shown is $\sim$14'$\times$12'. The position of FO Aqr is highlighted (arrow) and also the comparison star (C1).}
\end{minipage}
\end{figure}

All the images were reduced using \textit{IRAF} \cite{tod86} with typical calibration corrections of dark current and flatfield. Special \textit{IRAF} routines were used to align the sequences of images and aperture photometry \cite{ste87} was ostensibly used. We used the differential photometry technique to construct all the light curves showed in this work, where the instrumental magnitude of the interest object is discounted by the instrumental magnitude of one comparison star in the same field of view. Figure 10 shows the FOV gathered with OAUNI around FO Aqr. The frame corresponds to an 2.2 hours stacked sequence observed on 2016/08/02. The comparison star (C1 = TYC 5803-398-1, with \textit{R} = 10.49 mag~\cite{wen00}) used for differential photometry also is indicated. Col. (6) in Fig.~\ref{foaqrtab} shows the mean absolute calibrated \textit{R} magnitude for FO Aqr for each sequence. This calibration in  magnitude also was done using the comparison star. The shown magnitudes confirm that our observations were gathered during the 2016 FO Aqr low state (\textit{R} $\sim$ 15 mag).

\begin{figure}[t]
\begin{minipage}{12cm}
\includegraphics[width=12cm]{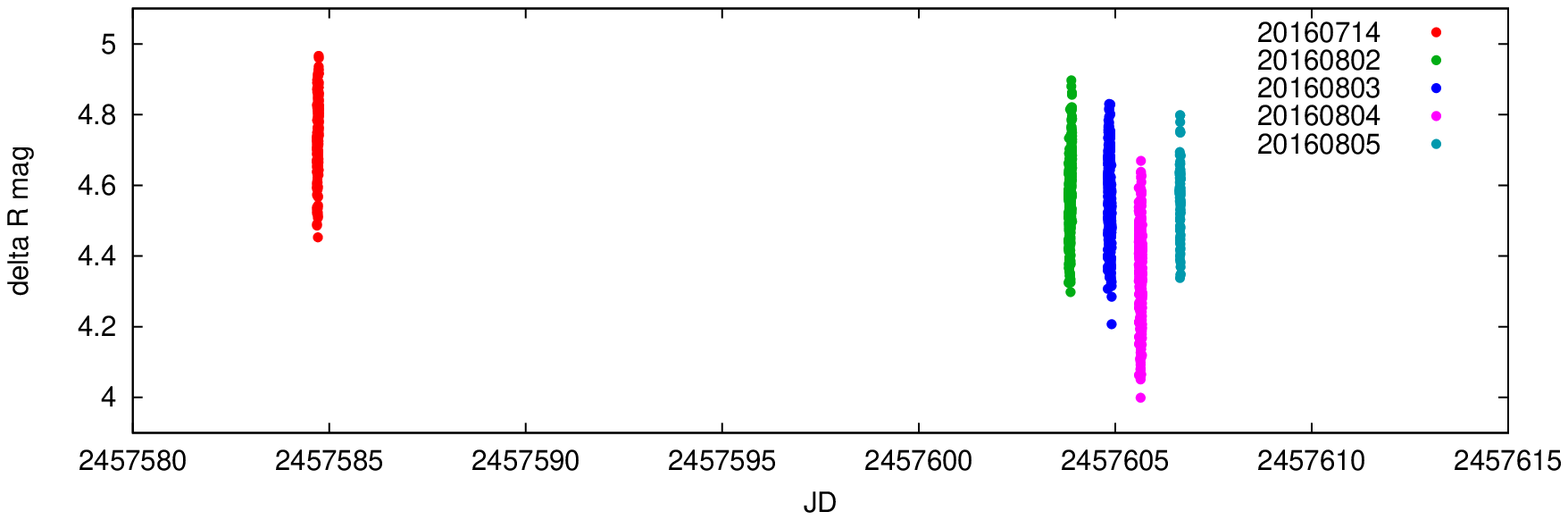}
\end{minipage}
\begin{minipage}{3.9cm}
\caption{\label{flc}FO Aqr light curves. The five-nights monitoring on 2016 campaign is shown.}
\end{minipage}
\begin{minipage}{12cm}
\includegraphics[width=12cm]{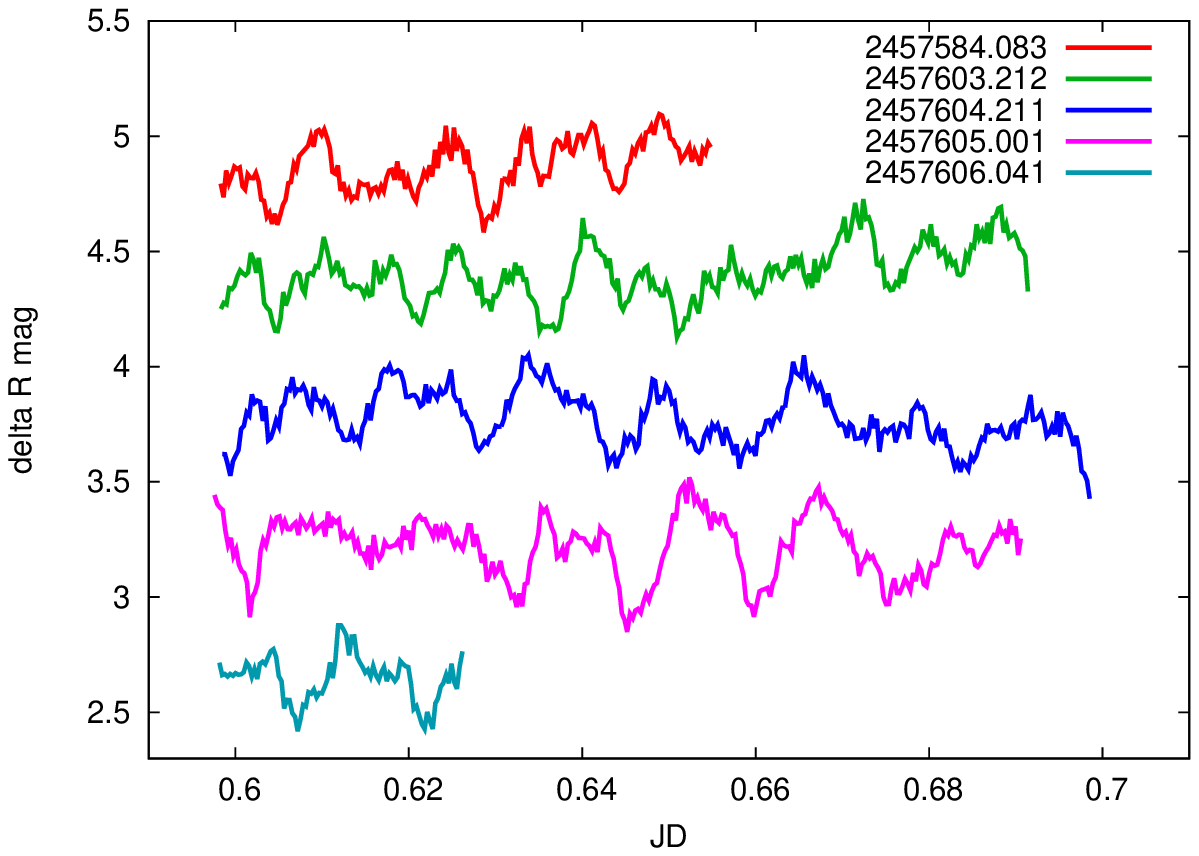}
\end{minipage}
\begin{minipage}{3.9cm}
\caption{\label{fdet} As Fig.~\ref{flc} including proper time (up right) and magnitude ($-$0.13, +0.17, +0.78, +1.15, and +1.92: from up to bottom, respectively) offsets in order to show in detail the stellar variability for FO Aqr.}
\end{minipage}
\end{figure}

\section{Analysis}
Figure~\ref{flc} shows the FO Aqr light curves observed during 2016 campaign. Figure~\ref{fdet} highlights the fast variability of the system in each night. Comparing with the typical mean error in our differential photometry ($<\sigma>_{phot}$ = 0.02 mag) by individual measurement, the variation scale of FO Aqr is at least one order of magnitude higher than $<\sigma>_{phot}$. In general, this condition is necessary to detect hidden periods in photometric time series. In addition, the light curves for the second, third and fourth nights present the longest uninterrupted time sequences (between 2.2 and 2.4 hrs).  Considering variability periods of tens of minutes as in FO Aqr, continuous times series of at least one order of magnitude higher are recommendable.

\begin{figure}
\center
\includegraphics[width=16cm]{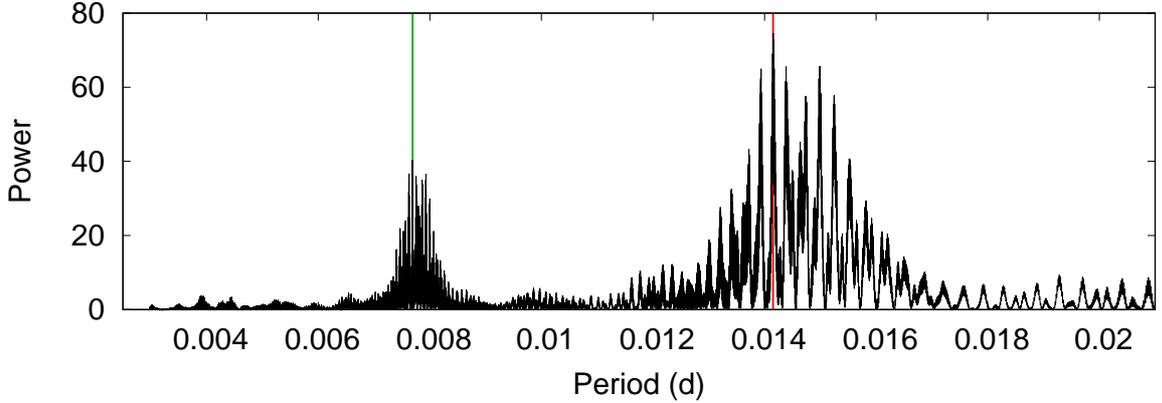}\hspace{2pc}%
\caption{\label{flomb}FO Aqr Lomb-Scargle periodogram. The two  main periods are shown, P$_{1}$~=~20.380~$\pm$~.003~min (red line) and P$_{2}$~=~11.076~$\pm$~.001~min (green line). The abscissa is in days.}
\end{figure}

\begin{figure}
\begin{minipage}{10cm}
\includegraphics[width=10cm]{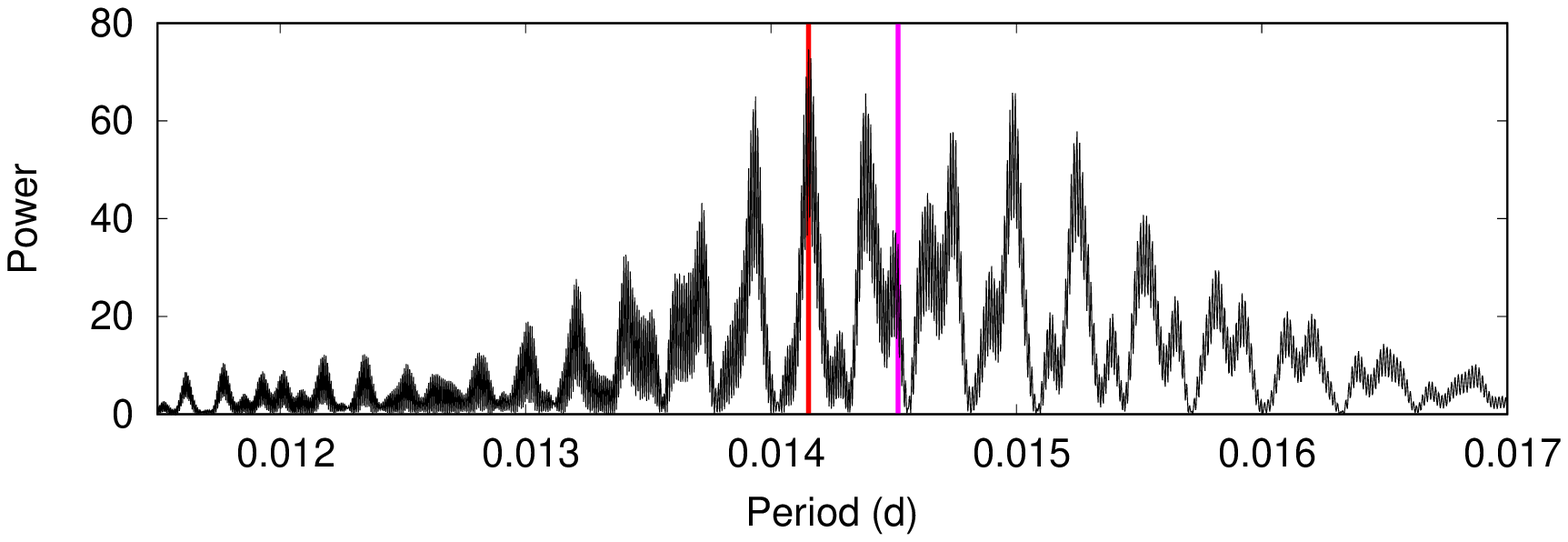}\hspace{2pc}%
\
\includegraphics[width=10cm]{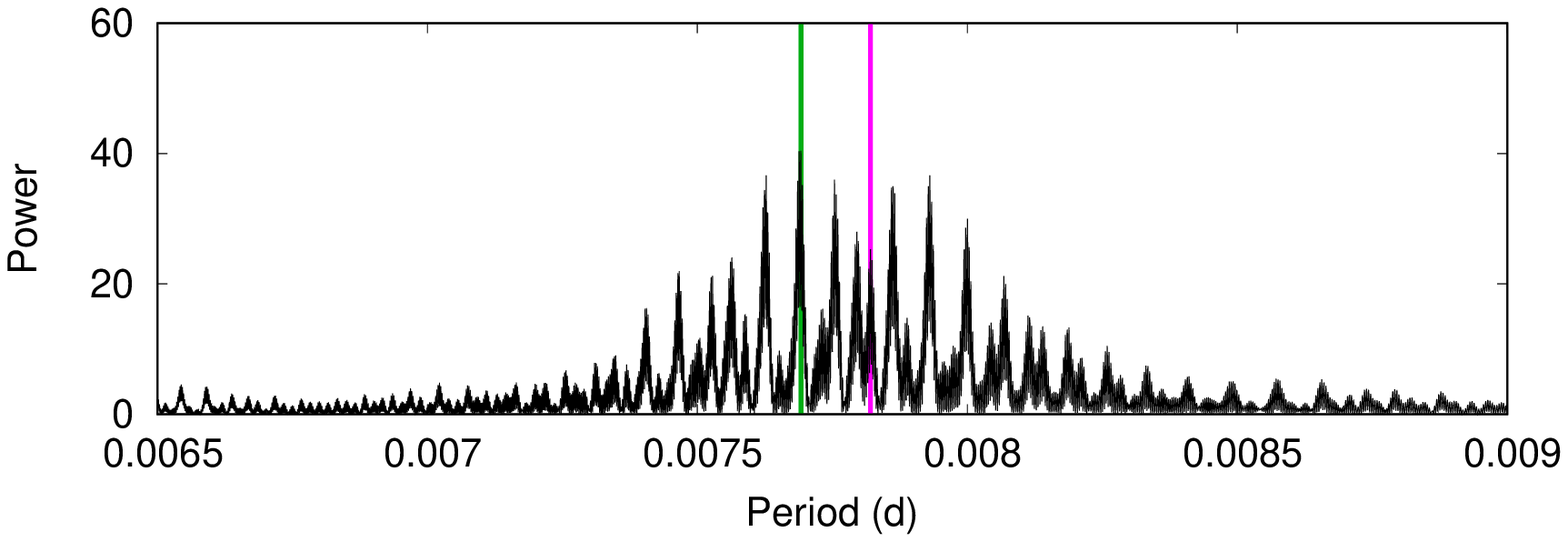}\hspace{2pc}%
\end{minipage}
\begin{minipage}[t]{5.9cm}
\caption{\label{fzcomp}Highlights of Fig.~\ref{flomb}. \textit{(Top)} P$_{1}$ (red line) compared with WD spin period (magenta line, \cite{ken16}). \textit{(Bottom)} P$_{2}$ (green line) compared with the one-half beat period (magenta line, \cite{ken16}).}
\end{minipage}
\end{figure}

Figure~\ref{flomb} shows the Lomb-Scargle periodogram \cite{lom76,sca82} for all our light curves of FO Aqr. For this purpose was used the period analysis software PERANSO \cite{pau16}. The time interval to look for periods was between 0.0025 to 0.025 days (or 3.5 to 26 min). This interval matches the two main known periods of FO Aqr which shown well defined peaks in Fig.~\ref{flomb}. The period with the higher power (P$_{1}$~=~20.380~$\pm$~.003~min) corresponds to the WD spin period usually observed in the high state~\cite{pat83}. The second peak (P$_{2}$~=~11.076~$\pm$~.001~min) corresponds to the one-half beat period which was detected very prominent at the beginning on July 2016 during the low state~\cite{lit16b}, just one week before our first observation. In order to compare different algorithms to construct the respective periodogram, the phase dispersion minimization (PDM) technique \cite{ste78} also was applied to the data and the results were very similar. 

\begin{figure}
\center
\includegraphics[width=16cm]{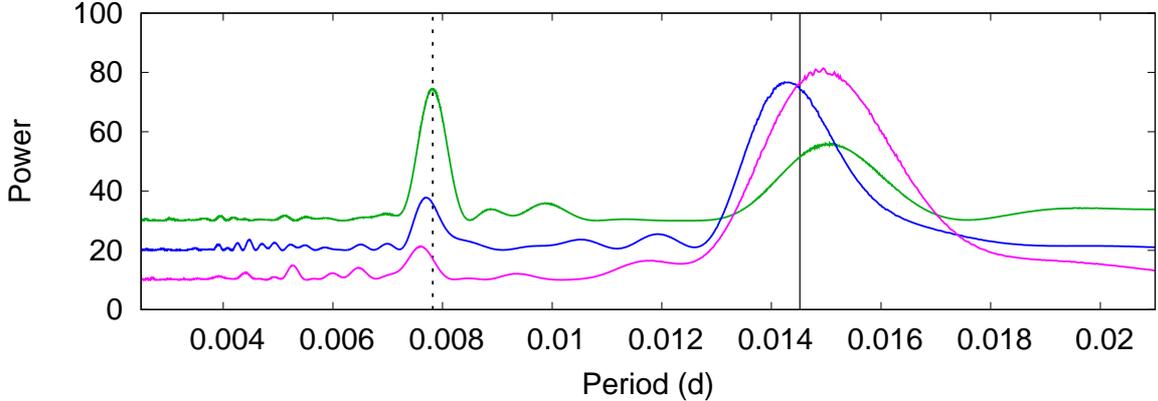}%
\caption{\label{perind}Individual periodograms for 20160802 (green), 20160803 (blue) and 20160804 (magenta) time series. The WD spin (black solid line) and the one-half beat (black dotted line) periods \cite{ken16} are shown to help the comparison. Note the excellent accuracy for P$_{2}$ in 20160802 data.}
\end{figure}

Figure~\ref{fzcomp} indicates that the periods P$_{1}$ and P$_{2}$ computed here are discrepant with \cite{ken16} in 31.5 and 11.2 seconds, respectively. %Eventually, these  accuracies can be representing a not precise time calibration process.%
 However, in our data P$_{1}$ is still the prominent period for FO Aqr in opposition to \cite{lit16a}. This suggests a possible intraday variation (or at scales of the orbital phase) of the prominent period during the low state. To check this possibility, individual periodograms were constructed for the available three largest photometric time series (see Figure~\ref{perind}). In addition, the interval of the orbital phase ($\phi_{int}$) was computed for each time series using the ephemerids from \cite{ken16} (Col. 7 in Table~\ref{foaqrtab}). The comparison shows that only on the 2016/08/02 data the one-half beat period is more prominent that the spin one. The next two days the pattern follows our general trend (Fig. \ref{flomb}). The 2016 August 02 result is consistent with \cite{lit16b} where this pattern was found on 2016 July 02, 06 and 07. These facts show evidence of a dependence of the beat period with the orbital phase of the system. In particular our 2016/08/02  pattern was observed at $\phi_{int}$ = 0.37$-$0.83 and with the best accuracy reached in our data. The discrepancy with respect to the one-half beat period in \cite{ken16} is only of 0.11 sec. Nevertheless, our time series one day after (2016/08/03) happens practically with the same $\phi_{int}$ (= 0.31$-$0.80) but with spin period higher. If the low state in FO Aqr is the result of a reduced accretion rate, the alternative prominence between P$_{1}$ and P$_{2}$ can also indicate important variations still at orbital period scales as suggested by \cite{lit16a} and also at the same phase interval.

\

\section{Conclusions}

The stellar variability of the intermediate polar FO Aqr was monitored on its 2016 low state using the OAUNI facility. The analysis of five light curves on 2016 July and August shows well defined periods associated with the white dwarf spin and the one-half beat periods. The WD spin period results to be the more prominent. Nevertheless, at least in one night the one-half beat modulation is higher with the better accuracy obtained for a given period in our data. Comparison of the orbital phase intervals where our light curves happen indicates important variations of the prominent period during a given orbital period on the low state.   

\ack
The author is grateful for the economic support from The World Academy of Sciences (TWAS), Rectorate and the Instituto General de Investigaci\'on (IGI) at UNI, and Concytec (Convenio 102-2015 Fondecyt). Special thanks to the Huancayo Observatory staff for the logistic support.

\end{document}